\documentclass[aps,prb,showpacs,superscriptaddress, preprint numbers,twocolumn]{revtex4-1}
\usepackage{rotating}
\usepackage{epstopdf}
\usepackage{color}
\usepackage{amsmath}
\usepackage{graphicx}
\usepackage{hyperref}
\usepackage{color}
\usepackage{amssymb}
\usepackage{dcolumn}
\usepackage{bm}
\usepackage{epsfig}
\usepackage{array}
\usepackage{subfigure}
\usepackage{upgreek}
\usepackage{wasysym}
\usepackage{slashed}
\usepackage[section]{placeins}
\usepackage{appendix}
\usepackage{subfigure}
\newcommand{\be}{\begin{equation}} 
\newcommand{\ee}{\end{equation}} 
\newcommand{\bea}{\begin{eqnarray}} 
\newcommand{\eea}{\end{eqnarray}} 
\newcommand{\bqa}{\begin{eqnarray}}
\newcommand{\eqa}{\end{eqnarray}}

\newcommand{\nn}{\nonumber \\}

\newcommand{\w}{\omega}

\newcommand{\figcomp}{
\begin{figure}[htb]
\centering
\hspace{0cm}
\subfigure[noonleline][]
{\label{fig: comp_moment_T10}\includegraphics[height=35mm,width=40mm]{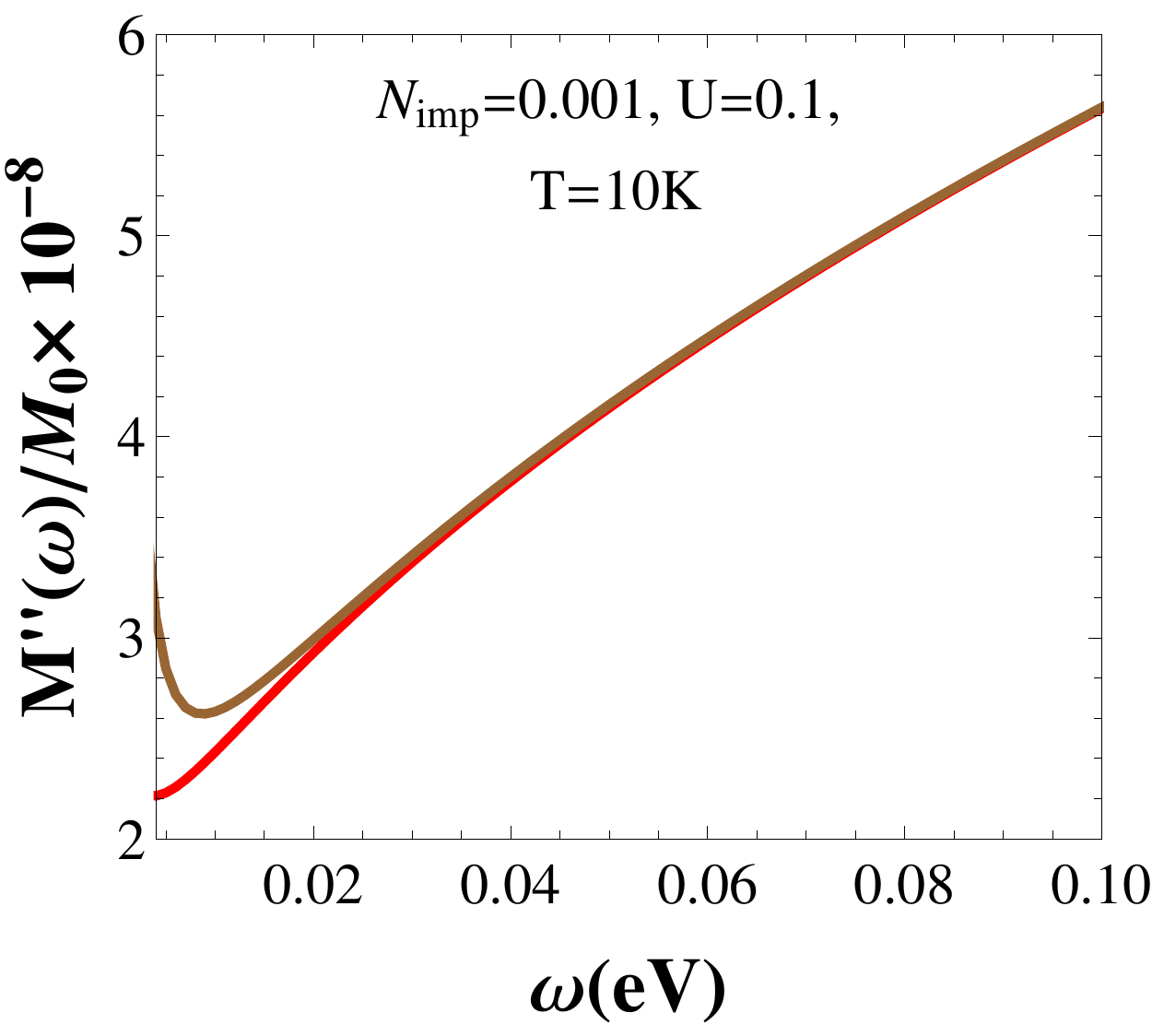}}
\hspace{0cm}
\subfigure[noonleline][]
{\label{fig: comp_moment_T200}\includegraphics[height=35mm,width=40mm]{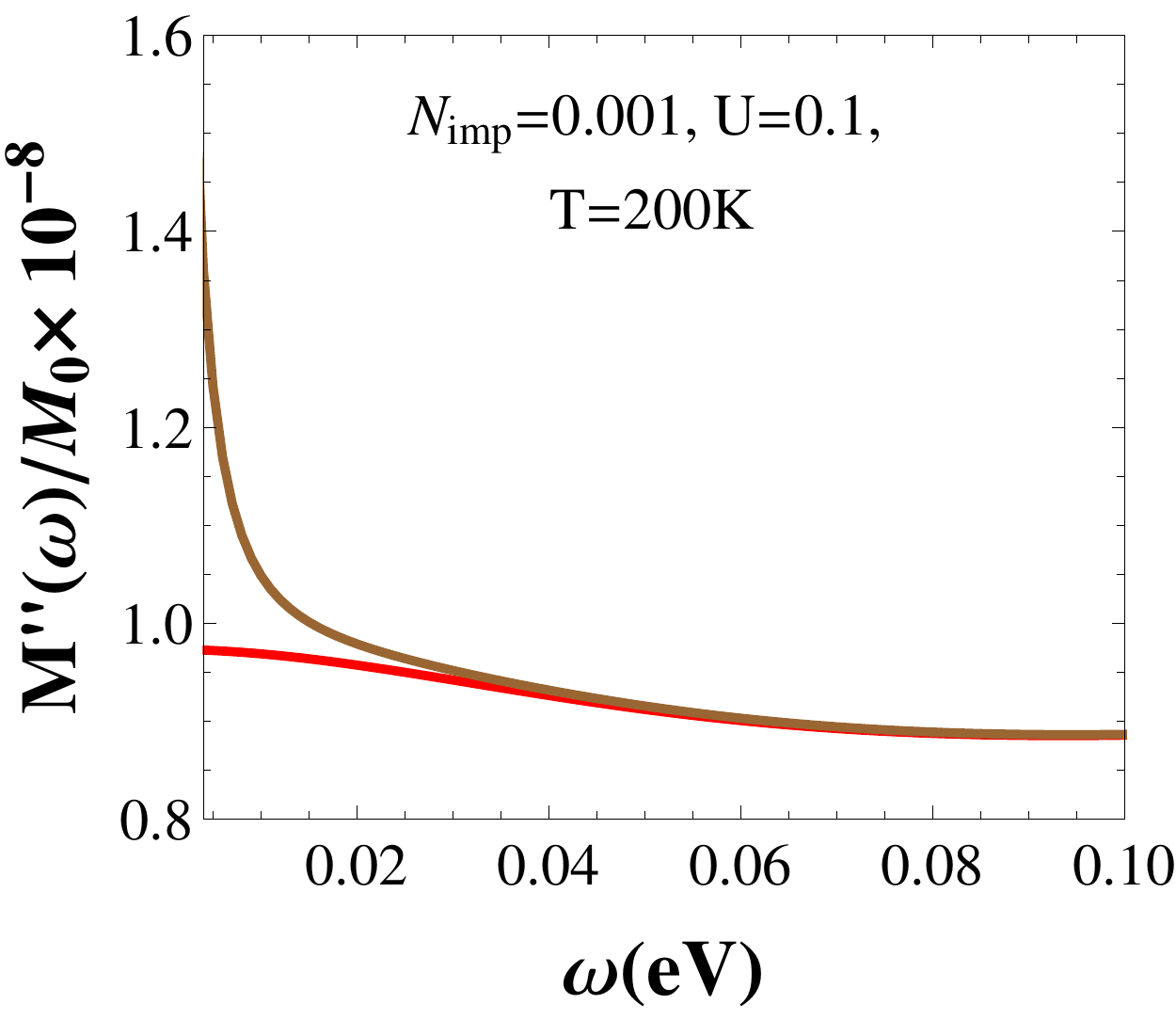}}
\caption{Plots of the imaginary part of normalized memory functions at different temperatures (a) at $T=10$K and (b) at $T=200$K. Here the red curve corresponds to the case with first moment only and the brown curve corresponds to the case where second moment also considered within the present moment expansion of the memory function. In both cases, there is nice agreement between the results from the two different approaches at high frequency regimes. However they differ significantly in the low frequency regime.}
\label{fig: comparison1}
\end{figure}
}
\newcommand{\figcompN}{
\begin{figure}[htb]
\centering
\hspace{0cm}
\subfigure[noonleline][]
{\label{fig: comp_moment_Nimpp01}\includegraphics[height=35mm,width=40mm]{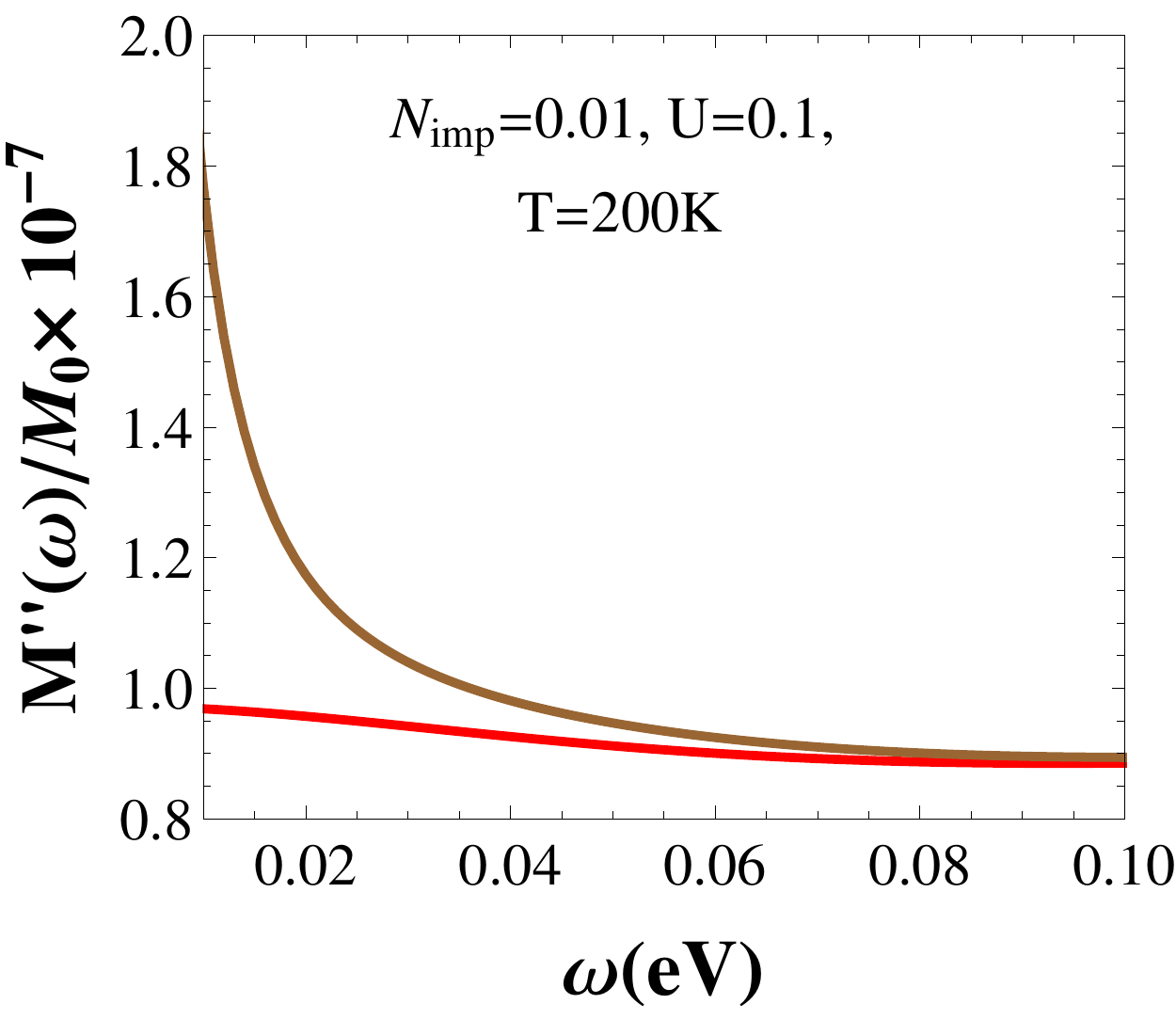}}
\hspace{0cm}
\subfigure[noonleline][]
{\label{fig: comp_moment_Nimpp04}\includegraphics[height=35mm,width=40mm]{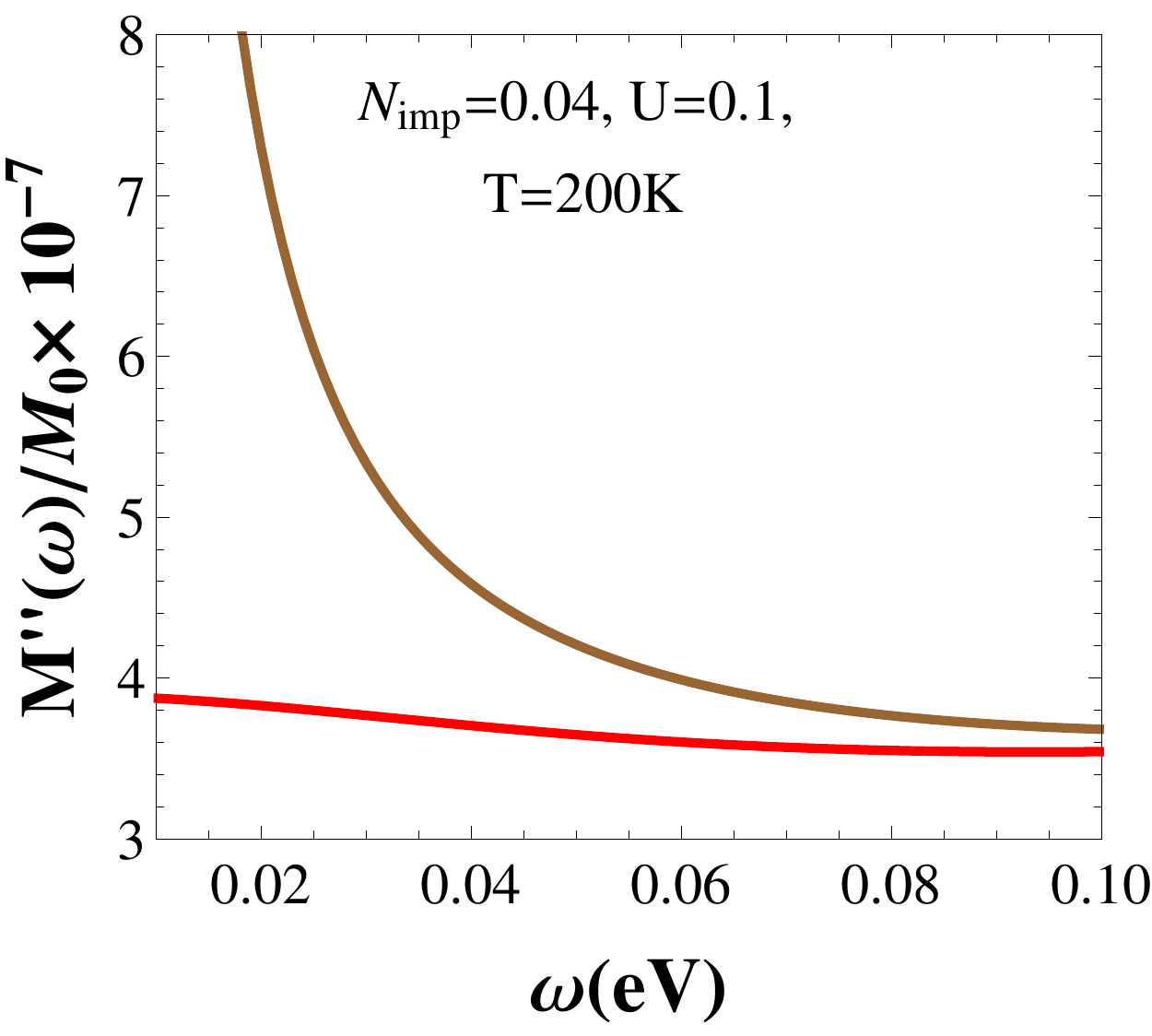}}
\caption{ Plots of the imaginary part of normalized memory functions at different impurity densities $N_{\text{imp}}$ (a) $0.01$ and (b) $0.04$. Here the red curve corresponds to the case with first moment only and the brown curve corresponds to the case where the second moment is also considered in the moment expansion. Here also a deviation occurs at low frequency regime as in the previous case. The increase in the impurity density enhances the magnitude of the memory function.}
\label{fig: comparison2}
\end{figure}
}
\newcommand{\figcompU}{
\begin{figure}[htb]
\centering
\hspace{0cm}
\subfigure[noonleline][]
{\label{fig: u1}\includegraphics[height=35mm,width=40mm]{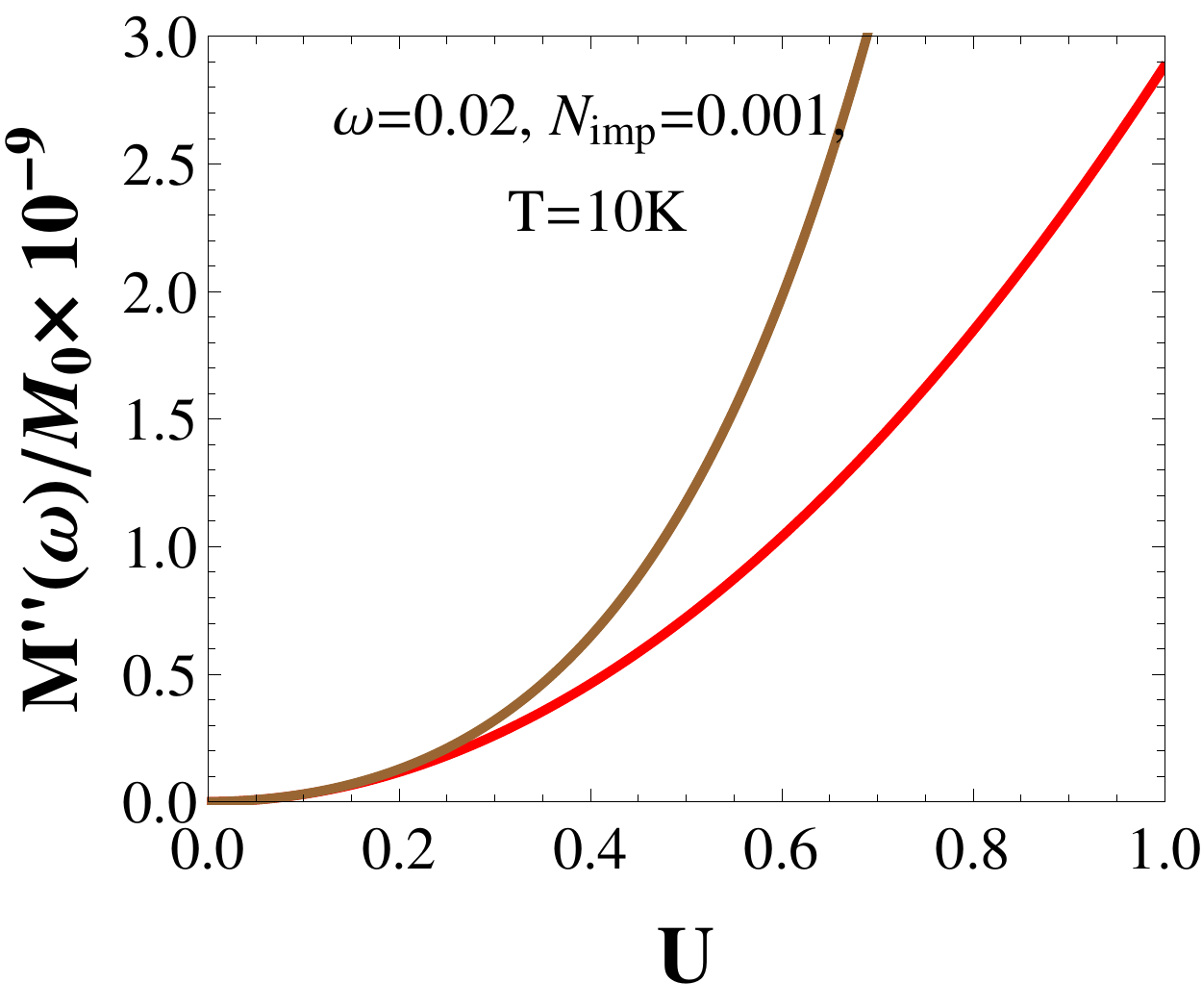}}
\hspace{0cm}
\subfigure[noonleline][]
{\label{fig: u2}\includegraphics[height=35mm,width=40mm]{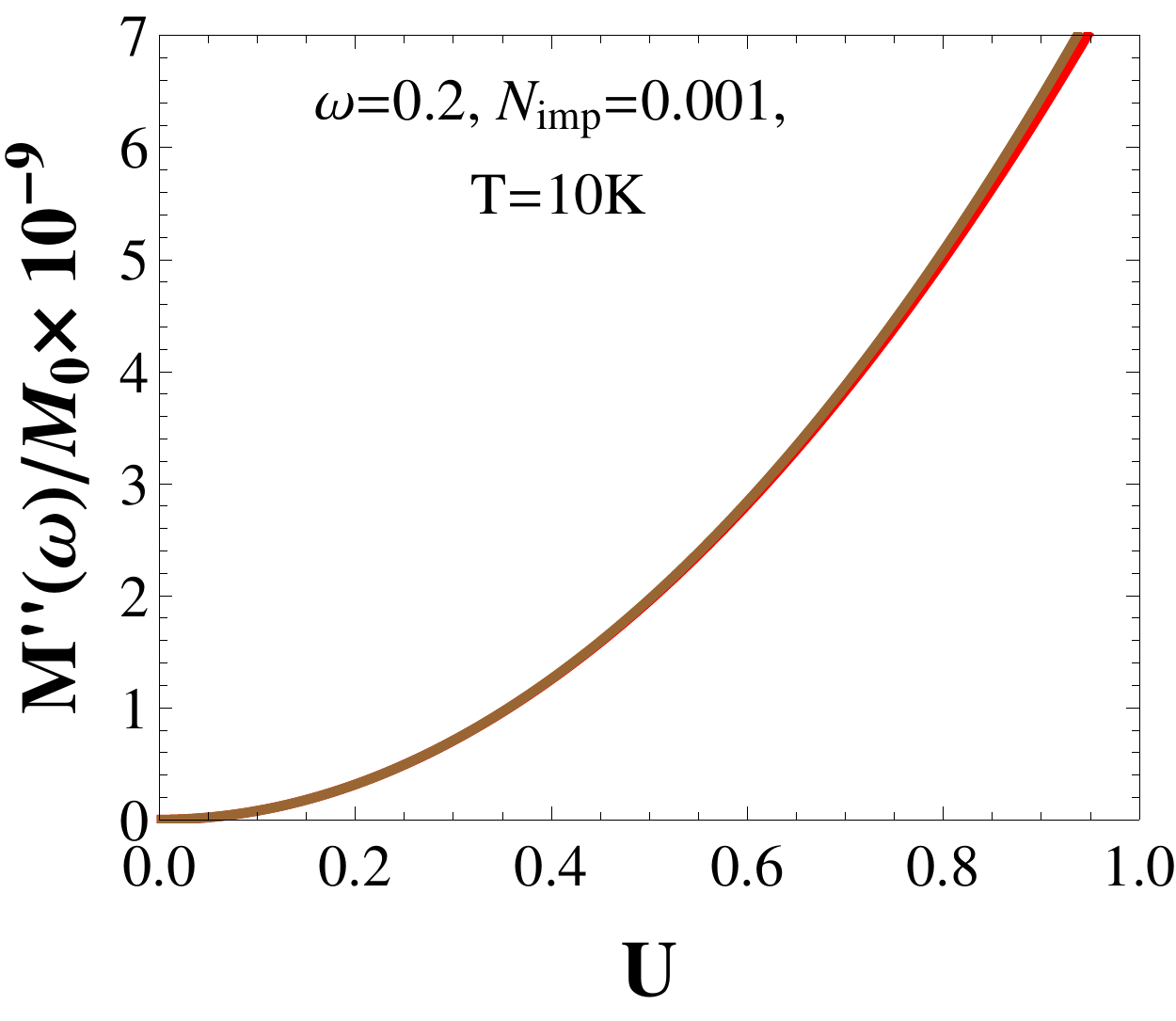}}
\caption{Variation of the scattering rates with interaction strength $U$ at different frequencies (a) $\omega=0.02$eV and (b) $0.2$eV. Here the red curve represents the scattering rate with the first moment only and the brown curve is with the inclusion of the second moment. It is observed that the deviation is more for higher interaction strength in the low frequency regime.}
\label{fig: comp_moment_interaction}
\end{figure}
}
\begin{document}
\title{Moment Expansion to the Memory Function for Generalized Drude Scattering rate}

\author{Pankaj Bhalla}
\email{pankajbhalla66@gmail.com}
\affiliation{Physical Research Laboratory, Navrangpura, Ahmedabad-380009 India.}
\affiliation{Indian Institute of Technology Gandhinagar-382355, India.}
\author{Nabyendu Das}
\affiliation{Physical Research Laboratory, Navrangpura, Ahmedabad-380009 India.}
\author{Navinder Singh}
\affiliation{Physical Research Laboratory, Navrangpura, Ahmedabad-380009 India.}
\date{\today}
\begin{abstract}
The memory function formalism is an important tool to evaluate the frequency dependent electronic conductivity. It is previously used within some approximations in the case of electrons interacting with various other degrees of freedom in metals with great success. However, one needs to go beyond those approximations as the interaction strengths  become stronger. In this work, we propose a systematic expansion of the memory function involving its various moments. We calculate the higher order contribution to the generalized Drude scattering rate in case of electron-impurity interactions. Further we compare our results with the results from previously studied lowest order calculations. We find larger contributions from the higher moments in the low frequency regime and also in the case of larger interaction strength.
\end{abstract}
\maketitle
\section{Introduction}
\label{sec: intro}
The study of frequency dependent conductivity or optical conductivity is very important to understand various interactions in the electronic systems \cite{basov_05, basov_11}.  In case of non-interacting electrons (neglecting coulomb interactions) colliding with ions, it can be cast in the simple Drude formula, where the optical conductivity $\sigma(\omega)$ is expressed as $\sigma(\omega)=\frac{\sigma_{0}}{1-i\omega\tau}$\cite{ashcroft_book}. Here $\sigma_{0}=\frac{ne^2\tau}{m}$ is the DC conductivity, where $n$ is the electron density, $m$ is the electron mass and $1/\tau$ refers to the scattering rate. Strictly speaking, the above Drude expression for optical conductivity is valid when $\omega << 1/\tau $. Thus we see that the frequency regime over which the Drude theory is valid depends on the smallness of the scattering rate $1/\tau$. The later increases with the increase of interaction strength and the validity regime shrinks. In presence of interactions, a modified form of the Drude conductivity with frequency dependent scattering rate is often used and the resulting expression is known as the generalized Drude conductivity \cite{timusk_03, puchkov_96, bhalla_14}. Within the linear response theory, the frequency dependent scattering rate ($1/\tau(\omega)$) is related to the current-current correlation which is equivalent to the two particle correlation functions\cite{mahan_book}. It captures the effects of different interactions within an electronic system.

The correlation functions can be calculated by several ways such as Mori's formalism\cite{mori_65}, within Pade approximation\cite{baker_75}, Ruelle response theory\cite{wouters_13}, generalized methods for recursion relations\cite{haydock_72, lee_82, lee_82a, hong_82}, etc. In general any formalism based on standard quantum many body perturbation theory, expresses two particle correlators in terms of single particle correlations\cite{mahan_book}. Thus the current-current correlator is expressed in terms of single particle correlators or single particle spectral function and the formalism depends on the existence of the quasi-particle. On the other hand the Mori-Zwanzig memory function formalism\cite{mori_65, zwanzig_61a, zwanzig_61} deals with the two particle correlators. It is based on the existence of few slow modes (e.g. conserved or nearly conserved electric current) related to certain conservation laws in the system. Hence the existence of quasiparticles is not a necessity here and this approach has wider range of applicability. The detailed discussions on its application in correlated electronic system can be found in a recent review by the present authors \cite{nabyendu_16}. In this method, the generalized scattering rate $1/\tau(\omega)$ can be expressed as an imaginary part of a memory function ($Im M(\omega)$). The later will be defined in the next section.

In literature, the memory function approach has been used in various systems, such as to study the molecular dynamics, thermodynamic properties, transport properties, etc\cite{gotze_71, gotze_72, arfi_92, plakida_96, plakida_97, vladimirov_12, forster_95, fulde_12, sega_03, prelovsek_04, sega_09, sega_06, sega_06a, maldague_77, grigolini_83, lucas_15, subir_15, subir_14, nabyendu_15, bhalla_15}. It becomes a method of choice in various strongly correlated electronic systems such as strange metal phase of the optimally doped cuprate superconductors where the very notion of the electronic quasiparticle breaks down \cite{subir_14, lucas_15}, but the translational invariance is present. In a generic electronic system there can be various slow modes such as the charge diffusion, the heat diffusion etc\cite{subir_14, lucas_15}. In the present study, we consider the electric current as the only relevant slow mode. We then systematically study the effects of other fast degrees of freedom on the current-current correlation within this formalism. In the present case, our main focus will be only on the role of electron-impurity interactions on the current-current correlation. The effects of the impurity interactions on the dynamical conductivity of a simple metal have been studied previously within the memory function in Ref.[\cite{gotze_71}] in detail. There authors  yield identical results for electrical conductivity with that of  the Boltzman's results \cite{ziman_book} in the dc limit. However the formalism is restricted to the lowest order in interaction strength and and needs corrections as the later increases.

With this motivation, we review the application of the memory function (MF) formalism in case of current-current correlation in metals and propose an expansion in terms of its various moments. Then we show that the previously studied G\"otze-W\"olfle\cite{gotze_72} formalism and similar other studies \cite{maldague_77, arfi_92, nabyendu_15, bhalla_14} are equivalent to the truncation our proposed moment expansion at the lowest order. We look for the case of higher interaction strength and calculate the contribution from the next order in the moment expansion.

This paper is organized as follows: In Sec.\ref{sec: MFF}, we present the memory function formalism for electrical conductivity. In Sec.\ref{sec: EQM}, the memory function is derived using equation of method approach. Then in Sec.\ref{sec: Firstmoment} the scattering rate has been calculated for impurity interactions with first moment expansion as done in literature. Then, we derived the second moment expansion of scattering rate and give the expression of scattering rate up-to second moment in  our expansion of the memory function in Sec.\ref{sec: higher}. In Sec.\ref{sec: results}, we compare our results with the former results. In Sec.\ref{sec: discussion}, we conclude with discussion. 
\section{Memory Function Formalism}
\label{sec: MFF}
The memory function method, also known as projection operator method is first introduced by 
Zwanzig\cite{zwanzig_61, zwanzig_61a} to study the time evolution of correlation functions. 
Later, the method was generalized by Mori\cite{mori_65} and the Laplace 
transform of an autocorrelation function was cast into a continued fraction form. In this section, we will review the mathematical description of the memory function formalism\cite{forster_95}. \\
Let us consider a system with a given Hamiltonian $H$ in which Liouville operator $\mathcal{L}$ is defined by its action on any operator $A$ as,
\begin{equation}
\mathcal{L}A = [H,A] = -i\frac{dA}{dt}.
\end{equation}
Here $A$ is an operator representing some observable and $[\cdots, \cdots]$ represents the commutator between two such operators and we use units in which $\hbar = 1$ and $k_{B} = 1$.
The above equation yields the time evolution of the operator as,
\begin{equation}
A(t)=e^{i\mathcal{L}t}A(0). 
\label{timeevolove} 
\end{equation}
To understand the dynamic property of certain observable in a many body systems, the time evolutions of related operators are needed. Let $A_{i}$ represents such operators. Their correlation is expressed in terms of the correlation function matrix $\mathcal{R}(t)$. The later,  in terms of its matrix elements is defined as,
\begin{equation}
R_{ij}(t) = \langle A_{i}(t) \vert A_{j}(0) \rangle.
\end{equation}
Here the inner product of such operators is defined as canonical ensemble average. Using the eqn.(\ref{timeevolove}) and performing the Laplace transform, the above equation can be expressed as,
\begin{equation}
R_{ij}(z) = \int_{0}^{\infty} dt e^{izt} \langle A_{i}(t) \vert A_{j}(0) \rangle   = \left\langle A_{i} \left\vert \frac{i}{z-\mathcal{L}} \right\vert A_{j} \right\rangle.
\label{correlation1}
\end{equation}
Here $z$ is a complex frequency and $z = \omega + i\eta$ with $\eta \rightarrow 0^+$. To express the correlation function in terms of the memory function, we introduce a projector operator $P$ which projects onto an operator $A$ and is defined as,
\begin{equation}
P=\sum_{i,j}\frac{ \vert A_{i} \rangle \langle A_{j} \vert }{\langle A_{i} \vert A_{j} \rangle}= \mathcal{I}-Q.
\end{equation}
Replacing the operator $\mathcal{L}$ by $\mathcal{L}(P+Q)$ in eqn.(\ref{correlation1}) and using the identity
\begin{equation}
\frac{1}{X+Y} = \frac{1}{X} - \frac{1}{X} Y \frac{1}{X+Y},
\end{equation}
the matrix elements of correlation function (eqn.(\ref{correlation1})) becomes,
\begin{equation}
R_{ij} = \left\langle A_{i} \left\vert \left\lbrace \frac{1}{z-\mathcal{L}Q} + \frac{1}{z-\mathcal{L}Q} \mathcal{L}P \frac{1}{z-\mathcal{L}} \right\rbrace \right\vert A_{j} \right\rangle.
\end{equation}
On simplification, the above expression can be rewritten as,
\begin{equation}
R_{ij} = \frac{1}{z} \chi_{ij} + \sum_{lm} \left\langle A_{i} \left\vert \frac{1}{z-\mathcal{L}Q} \mathcal{L} \right\vert A_{l} \right\rangle \chi_{lm}^{-1} R_{mj},
\end{equation}
where $\chi_{ij}=\langle A_{i} \vert A_{j} \rangle $. In matrix notation, this can be written as,
\begin{equation}
(z\mathcal{I}-\mathcal{K}\chi^{-1})\mathcal{R}=\chi.
\end{equation}
Here the elements of matrix $\mathcal{K}$ are defined as,
\begin{eqnarray} \nonumber
K_{il} &=&  \left\langle A_{i} \left\vert \frac{z}{z-\mathcal{L}Q}\mathcal{L} \right\vert A_{l} \right\rangle \\
&=& \langle  A_{i} \vert \mathcal{L} \vert A_{l} \rangle + \left\langle A_{i} \left\vert \mathcal{L}Q \frac{1}{z-\mathcal{L}Q} \mathcal{L} \right\vert A_{l} \right\rangle. 
\end{eqnarray}
The first part of the right hand side of the above equation is known as frequency matrix and is defined as,
\begin{equation}
\mathcal{L}_{il}=\langle A_{i} \vert\mathcal{L} \vert A_{l} \rangle.
\end{equation}
The other part is known as memory matrix and is defined as follows,
\begin{equation}
M_{il}(z) = \left\langle A_{i} \left\vert \mathcal{L}Q \frac{1}{z-\mathcal{L}Q}\mathcal{L} \right\vert A_{l} \right\rangle.
\end{equation}
Using the fact $Q^{2}=Q$,  the above expression can be written in a symmetric form  as,
\begin{equation}
M(z)= \left\langle A\left\vert \mathcal{L}Q \frac{1}{z-Q\mathcal{L}Q} Q\mathcal{L} \right\vert B \right\rangle.
\end{equation}
Now, on applying the Liouvilian operator 
on both the operators $A$ and $B$, the above equation reduces to
\begin{equation}
M(z)= \left\langle \dot{A}\left\vert Q \frac{1}{z-Q\mathcal{L}Q} Q \right\vert \dot{B} \right\rangle.
\end{equation}
As we focus on the electrical conductivity and thus our concern is the current-current correlation. Hence, we replace both $A$ and $B$ operators by the current operator $J$. Thus the desired memory function for the electrical conductivity becomes,
\begin{equation}
M(z)= \left\langle \dot{J}\left\vert Q \frac{1}{z-Q \mathcal{L} Q} Q \right\vert \dot{J} \right\rangle.
\end{equation}
On expanding $M(z)$ in series expansion, we have
\begin{eqnarray}  
M(z)&=&\frac{1}{z}  \left\langle \dot{J}\left\vert Q \left(1+ \frac{1}{z} Q \mathcal{L} Q \right. \right. \right. \nn
&& \left. \left. \left.+ \frac{1}{z^{2}} Q \mathcal{L} QQ \mathcal{L} Q + \cdots \right) Q\right\vert \dot{J}\right\rangle.
\end{eqnarray}
Using the fact that $QQ=Q^{2}=(1-P)^{2}= Q$ and $\langle J\vert \dot{J}\rangle$, $\langle \dot{J}\vert \ddot{J}\rangle$ $=0$ (proved in appendix \ref{app: A}), 
the memory function in series expansion can be written as
\begin{eqnarray}  
M(z)&=&\frac{1}{z} \langle \dot{J} \vert  \dot{J} \rangle  + \frac{1}{z^{3}}
\langle  \ddot{J} \vert  \ddot{J} \rangle +\cdots+\frac{1}{z^{2n-1}} \langle \overset{n}{J} \vert  \overset{n}{J}  \rangle.
\label{eqm6}
\end{eqnarray}
Here $\overset{n}{J}$ represents the $n^{\text{th}}$ time derivative of the current operator. This expression represents the high frequency expansion of the memory function interms of the equal time autocorrelation function. With this motivation, we will derive a similar expression for the memory function by an alternative way in next section.


\section{Equation of motion Method}
\label{sec: EQM}
In an alternative way, the memory function can also be calculated using the equation of motion method (EQM) as follows. Let us start with the expression for
response function within the linear response theory by Kubo\cite{kubo_57, kadanoff_63, zubarev_60}, which is given as,
\begin{equation}
\chi_{AB}(z)=\langle\langle A ; B \rangle\rangle_{z}= -i\int_{0}^{\infty} e^{izt} \langle [A(t), B(0)]\rangle dt .
\label{chi1}
\end{equation}
Here $A$ and $B$ are two operators and correspond to two physical variables, $[A, B]$ denotes their commutator and the inner $\langle \cdots \rangle$ represents statistical 
ensemble average at temperature $T$. The outer $\langle \cdots \rangle$ represents the Laplace transform at a complex frequency $z$. Using the equation of motion, $\langle\langle A ; 
B \rangle\rangle_{z}$ can be written as,
\begin{equation}
z\langle\langle A ; B \rangle\rangle_{z} = \langle [A, B] \rangle + \langle\langle [A,H] ; B \rangle\rangle_{z}.
\label{eqm1}
\end{equation}
Here $H$ is the total Hamiltonian of the system. According to the Heisenberg equation of motion, an operator evolves as, 
\begin{equation}
i\frac{dA}{dt}= i\dot{A} = [A,H].
\end{equation}
Using the above expression, the eqn.(\ref{eqm1}) can be expressed as,
\begin{equation}
z\langle\langle A ; B \rangle\rangle_{z} = \langle [A, B] \rangle +i \langle\langle \dot{A} ; B \rangle\rangle_{z}.
\end{equation}
In the present case, we are interested in current-current correlation function. Hence, we replace both $A$ and $B$ by current operator $J$. 
Thus, the above equation becomes
\begin{equation}
z\langle\langle J ; J \rangle\rangle_{z} = \langle [J, J] \rangle +i \langle\langle \dot{J} ; J \rangle\rangle_{z}.
\end{equation}
As the commutator $[J,J] = 0$, the above equation reduces to
\begin{equation}
z\langle\langle J ; J \rangle\rangle_{z} = i \langle\langle \dot{J} ; J \rangle\rangle_{z}.
\label{eqm2}
\end{equation}
Again from the equation of motion (using eqn.\ref{eqm1}),
\begin{equation}
z\langle\langle \dot{J} ; J \rangle\rangle_{z} = \langle[\dot{J},J]\rangle + i\langle\langle \dot{J} ; \dot{J} \rangle\rangle_{z}.
\end{equation}
For $z=0$, $\langle[\dot{J},J]\rangle = -i\langle\langle \dot{J} ; \dot{J} \rangle\rangle_{z=0}$. Using these, the eqn.(\ref{eqm2}) can be written as,
\begin{equation}
z\langle\langle J ; J \rangle\rangle_{z} = \frac{1}{z}\left(\langle\langle \dot{J} ; \dot{J} \rangle\rangle_{z=0} - 
\langle\langle \dot{J} ; \dot{J} \rangle\rangle_{z}\right).
\label{eqm3}
\end{equation}
This expression is used in the well cited work by G\"otze and W\"olfle\cite{gotze_72}  to evaluate the memory function for electrons in metal with various interactions. However instead considering the above expression and evaluating $
\langle\langle \dot{J} ; \dot{J} \rangle\rangle_{z}$ perturbatively, we can opt for a  higher moment expansion as follows. We apply EQM method again to evaluate the correlation 
function $\langle\langle J;J\rangle\rangle$ in terms of the correlations involving higher time derivatives of $\dot{J}$. Thus in order 
to express in next moment i.e. second moment, we use the EQM for $\langle\langle \dot{J} ; \dot{J} \rangle\rangle_{z}$,  and obtain, 
\begin{equation}
z\langle\langle \dot{J} ; \dot{J} \rangle\rangle_{z} = \langle[\dot{J},\dot{J}]\rangle + \langle\langle [\dot{J},H] ; 
\dot{J} \rangle\rangle_{z}.
\end{equation}
Using $\langle[\dot{J},\dot{J}]\rangle = 0$ and $z\langle\langle [\dot{J},H],\dot{J}\rangle\rangle = \langle\langle 
\ddot{J};\ddot{J}\rangle\rangle_{z=0} - \langle\langle \ddot{J};\ddot{J}\rangle\rangle_{z}$, the above equation can be written as
\begin{equation}
z\langle\langle \dot{J} ; \dot{J} \rangle\rangle_{z} = -\frac{1}{z}\left( \langle\langle \ddot{J};\ddot{J}\rangle\rangle_{z=0} - 
\langle\langle \ddot{J};\ddot{J}\rangle\rangle_{z} \right). 
\end{equation}
Substitute this equation in eqn.(\ref{eqm3}), we have
\begin{eqnarray} \nonumber
z\langle\langle J ; J \rangle\rangle_{z} &=& \frac{1}{z} \langle\langle \dot{J} ; \dot{J} \rangle\rangle_{z=0} \\
&& + \frac{1}{z^{3}}\left( \langle\langle \ddot{J};\ddot{J}\rangle\rangle_{z=0} - \langle\langle \ddot{J};\ddot{J}
\rangle\rangle_{z} \right).
\end{eqnarray}
Thus the expression for the response function becomes,
\begin{equation} 
z\chi(z) = \frac{1}{z} \langle\langle \dot{J} ; \dot{J} \rangle\rangle_{z=0}  + \frac{1}{z^{3}}\left( \langle\langle 
\ddot{J};\ddot{J}\rangle\rangle_{z=0} - \langle\langle \ddot{J};\ddot{J}\rangle\rangle_{z} \right).
\label{eqm4}
\end{equation}
By applying EQM again and again, we can obtain a series expansion for $z\chi(z)$ as,
\begin{eqnarray} \nonumber
z\chi(z) &=& \frac{1}{z} \langle\langle \dot{J} ; \dot{J} \rangle\rangle_{z=0} + \frac{1}{z^{3}} \langle\langle \ddot{J};
\ddot{J}\rangle\rangle_{z=0}- \cdots\\
&&  + \frac{1}{z^{2n-1}} \langle\langle \overset{n}{J};\overset{n}{J}\rangle\rangle_{z=0}- \frac{1}{z^{2n-1}} \langle\langle 
\overset{n}{J};\overset{n}{J}\rangle\rangle_{z}.
\label{eqm5} 
\end{eqnarray}
In Ref.[\cite{gotze_72}], it is shown that $\chi(z)$ is related to the memory function as
\begin{equation}
M(z)=z\frac{\chi(z)}{\chi_{0}-\chi(z)},
\end{equation}
where $\chi_{0}$ represents the static correlation function ($= N_{e}/m$, where $N_{e}$ corresponds to electron density).
Here $M(z)$ is the complex memory function, which upon analytic continuation, can be written as a function of real frequency as,
\begin{equation}
M(\omega \pm i0) = M'(\omega) \pm M''(\omega),
\end{equation}
where $ M'(\omega)$ and $ M''(\omega)$ are real and imaginary part of the memory function and satisfies the symmetry properties  $ M'(\omega) = - M'(-\omega)$ and $ M''(\omega) = M''(-\omega)$ \cite{gotze_72}.\\
An approximate form of the memory function can be obtained by assuming that $\chi(z)/\chi_{0}$ is smaller than one.  Within this appoximation, the  expression for the memory function becomes,
\begin{equation}
M(z)=\frac{z\chi(z)}{\chi_{0}} \left( 1+ \frac{\chi(z)}{\chi_{0}} - \cdots \right).
\end{equation}
Keeping only the leading order term, the memory function can be expressed as
\begin{equation}
M(z) = z \frac{\chi(z)}{\chi_{0}}.
\label{mem1}
\end{equation}
This expression is valid under the approximation discussed before and works well in high frequency regime and shows valid/invalid results in low frequency regime depending upon the parameters chosen to calculate the $\chi(z)$. The more details of its validity are discussed in our recent work\cite{bhalla_15}.\\
Using eqn.(\ref{eqm5}), the memory function to general order can be written as,
\begin{eqnarray} \nonumber
M(z)&=&\frac{1}{\chi_{0}} \left(\frac{1}{z} \langle\langle \dot{J} ; \dot{J} \rangle\rangle_{z=0} + \frac{1}{z^{3}} 
\langle\langle \ddot{J};\ddot{J}\rangle\rangle_{z=0}+ \cdots \right.\\ \nonumber
&& \left. \cdots + \frac{1}{z^{2n-1}} \langle\langle \overset{n}{J};\overset{n}{J}\rangle\rangle_{z=0}- \frac{1}{z^{2n-1}} 
\langle\langle \overset{n}{J};\overset{n}{J}\rangle\rangle_{z} \right).\\
\label{mem2}
\end{eqnarray}
This is an expression of the complex memory function which is equivalent to the eqn.(\ref{eqm6}), but under a restrictive condition $\chi(z) << \chi_{0}$ \cite{bhalla_15}.  Here we see that instead of limiting at a perturbative calulation of $\dot{J}-\dot{J}$ correlation, we can include correlations involving higher order  time derivatives of $\dot{J}$. Since the correlations with higher order time derivatives involves higher order corrections in interaction strength to the scattering rate. We will use this expression with $n=2$, to evaluate the scattering rate due to the impurity interactions in later sections and will see how the result differs from that of the previously studied lower order corrections.
\section{Case of electron-impurity scattering}
\label{sec: Firstmoment}
In this section, we review the work discussed in Ref.[\cite{gotze_72}] to calculate the memory function 
for impurity interactions. We consider a metal where degenerate electrons are interacting with impurities. In this case,  the Hamiltonian is described as
\begin{equation}
H= H_{0} + H_{\text{imp}}.
\end{equation}
Here $H_{0}$ is the unperturbed Hamiltonian and in second quantized notation can be written as\cite{mahan_book}
\begin{equation}
H_{0}= \sum_{\textbf{p}} \epsilon_{\textbf{p}} c^{\dagger}_{\textbf{p}} c_{\textbf{p}}.
\label{unperturbed}
\end{equation}
Here $c^{\dagger}_{\textbf{p}}$ and $c_{\textbf{p}}$ are electron creation and annihilation operators respectively and $\epsilon_{p}$ is the energy of free electrons with momenta $p$. The other part of Hamiltonian describes the electron-impurity interaction and is given as,
\begin{equation}
H_{\text{imp}}= \frac{1}{N} \sum_{j=1}^{N_{\text{imp}}} \sum_{\textbf{k},\textbf{k}',\sigma} \langle \textbf{k} 
\vert U^{j} \vert \textbf{k}' \rangle c^{\dagger}_{\textbf{k},\sigma} c_{\textbf{k}' \sigma},
\label{imp}
\end{equation} 
where $N$ represents the number of lattice cells, $N_{\text{imp}}$ corresponds to number of impurity sites and 
$U^{j}$ is the scattering potential from $j^{\text{th}}$ impurity.\\
Computation of the memory function  in Ref.[\cite{gotze_72}] is restricted to the first moment only. First we discuss it.  Truncating at the first order, the memory function can be written as,
\begin{equation}
M(z)=\frac{1}{z \chi_{0}} \left(\langle\langle \dot{J} ; \dot{J} \rangle\rangle_{z=0} - \langle\langle \ddot{J};\ddot{J}\rangle\rangle_{z} \right).
\end{equation}
To evaluate the above expression, let us first calculate $\dot{J}$. It s defined as,
\begin{eqnarray} 
\dot{J} &=& -i[J, H] = -i\left([J,H_{0}]+ [J, H_{\text{imp}}]\right).
\end{eqnarray} 
As $[J,H_{0}] = 0$, thus $\dot{J}= -i[J,H_{\text{imp}}]$. Using eqn.(\ref{imp}) and the defining  the current operator 
$J=\sum_{\textbf{k}} v_{x}(\textbf{k}) c^{\dagger}_{\textbf{k}} c_{\textbf{k}}$, where $v_{x}$ is the  x-component of velocity, the time derivative of $J$ can be written  as,
\begin{equation}
\dot{J}= -\frac{i}{N} \sum_{j,\textbf{k},\textbf{k}'} \langle \textbf{k} \vert U^{j} \vert \textbf{k}'\rangle 
\left( v_{x}(\textbf{k})- v_{x}(\textbf{k}'\right) c^{\dagger}_{\textbf{k}} c_{\textbf{k}'}.
\label{firstj}
\end{equation}
With the above expression, the correlator $\langle\langle \dot{J};\dot{J} \rangle\rangle$ becomes
\begin{eqnarray} \nonumber
\langle\langle \dot{J};\dot{J} \rangle\rangle_{z} &=& - \frac{1}{N^{2}} \sum_{j,\textbf{k},\textbf{k}'}
\sum_{i,\textbf{p},\textbf{p}'}  \langle \textbf{k} \vert U^{j} \vert \textbf{k}'\rangle \langle \textbf{p} 
\vert U^{i} \vert \textbf{p}'\rangle \\ \nonumber
&& \left( v_{x}(\textbf{k})- v_{x}(\textbf{k}')\right) \left( v_{x}(\textbf{p})- v_{x}(\textbf{p}')\right) \\
&& \langle\langle c^{\dagger}_{\textbf{k}} c_{\textbf{k}'};c^{\dagger}_{\textbf{p}} c_{\textbf{p}'} \rangle\rangle.
\label{firstmomemt55}
\end{eqnarray}
Using the definition of the correlator as defined in  eqn.(\ref{chi1}), $\langle\langle c^{\dagger}_{\textbf{k}} c_{\textbf{k}'};c^{\dagger}_{\textbf{p}} 
c_{\textbf{p}'} \rangle\rangle$ after doing time integration and thermal average by using $c_{\textbf{k}}(t) = c_{\textbf{k}} e^{i\epsilon_{\textbf{k}}t}$, we get,
\begin{equation}
-\frac{1}{z+\epsilon_{\textbf{k}}-\epsilon_{\textbf{k}'}} \left( f(\textbf{k})-f(\textbf{k}') \right) 
\delta_{\textbf{p}',\textbf{k}} \delta_{\textbf{p},\textbf{k}'}.
\end{equation} 
We consider the above expression and also the case of dilute impurity and neglecting the interference terms, thus substitute $i=j$ in eqn.(\ref{firstmomemt55}). Performing the summation over impurity sites which contributes 
$N_{\text{imp}}$, we have
\begin{eqnarray} \nonumber
\langle\langle \dot{J};\dot{J} \rangle\rangle_{z} &=& 2 \frac{N_{\text{imp}}}{N^{2}} \sum_{\textbf{k},\textbf{k}'} 
\vert \langle \textbf{k} \vert U \vert \textbf{k}'\rangle \vert ^{2} \left( v_{x}(\textbf{k})- v_{x}(\textbf{k}')\right)^{2} \\
&&  \times\frac{f(\textbf{k})-f(\textbf{k}')}{z+\epsilon_{\textbf{k}}-\epsilon_{\textbf{k}'}}.
\label{ijcase}
\end{eqnarray}
Here factor $2$ is due to the spin degeneracy. After simplification cosidering isotropic  free electron case and writing $\textbf{v}=\textbf{k}/m$,
\begin{eqnarray} \nonumber
\langle\langle \dot{J};\dot{J} \rangle\rangle_{z} &=& \frac{2}{3} \frac{N_{\text{imp}}}{m^{2}N^{2}} 
\sum_{\textbf{k},\textbf{k}'} \vert \langle \textbf{k} \vert U \vert \textbf{k}'\rangle \vert ^{2} 
\left(\textbf{k}- \textbf{k}'\right)^{2} \\
&& \times \frac{f(\textbf{k})-f(\textbf{k}')}{z+\epsilon_{\textbf{k}}-\epsilon_{\textbf{k}'}}.
\label{jdot2}
\end{eqnarray}
On substituting the above equation in eqn.(\ref{eqm3}) and  using the eqn.(\ref{mem1}),  followed by analytic continuation, i.e. $z \rightarrow \omega + i\eta$, $\eta
 \rightarrow 0^+$, the imaginary part of the memory function becomes,
\begin{eqnarray} \nonumber
M''(\omega) &=& \frac{2\pi}{3N^{2}} \frac{N_{\text{imp}}}{m N_{e}\omega} \sum_{\textbf{k}, \textbf{k}'} \vert 
\langle \textbf{k} \vert U \vert \textbf{k}'\rangle \vert ^{2} \left(\textbf{k}- \textbf{k}'\right)^{2} \\
&& \times \left(f(\textbf{k})-f(\textbf{k}')\right) \delta\left( \omega+\epsilon_{\textbf{k}}-\epsilon_{\textbf{k}'}  \right).
\end{eqnarray}
Under the assumption that $U$ is independent of momentum, i.e. for point like impurities \cite{bennemann_book, nabyendu_15a}
the expression further reduces to,
\begin{eqnarray} \nonumber
M''(\omega) &=& \frac{2\pi}{3N^{2}} \frac{N_{\text{imp}}U^{2}}{m N_{e}\omega} \sum_{\textbf{k}, \textbf{k}'} 
\left(\textbf{k}- \textbf{k}'\right)^{2} \\
&& \times \left(f(\textbf{k})-f(\textbf{k}')\right) \delta\left( \omega+\epsilon_{\textbf{k}}-\epsilon_{\textbf{k}'}  \right).
\end{eqnarray}
Converting the summation over momentum indices to the energy integrals and 
performing one integral involving the  delta function, the equation further reduces to
\begin{eqnarray} \nonumber
M''(\omega) &=& \frac{2}{3} \frac{N_{\text{imp}}}{N_{e}} \frac{U^{2}m^{3}}{\pi^{3}\omega} \int_{0}^{\infty} 
d\epsilon \sqrt{\epsilon(\epsilon+\omega)} \\
&& (2\epsilon + \omega) \left(f(\epsilon)-f(\epsilon')\right) .
\label{gwmem}
\end{eqnarray}
This is an expression of imaginary part of the memory function or the scattering rate of the electronic quasiparticles due to the electron-impurity 
interactions. Here for simplicity we replace $\epsilon_{\textbf{k}}$ and $\epsilon_{\textbf{k}'}$ by $\epsilon$ and $\epsilon'$ respectively in rest of the calculation. According to our proposed expansion, this result is equivalent to restrict the eqn.(\ref{mem2}) at $n=1$ followed by a perturbative evaluation of the $\dot{J}-\dot{J}$ correlation. In the next section we will perform a perturbative calculation at higher order and will show that this approximation has limited validity.
\section{The MF with a higher order moment}
\label{sec: higher}
The memory function with higher order moment can be calculated within the moment expansion proposed by us using eqn.(\ref{eqm5}). One can obtain more exact result by including higher order moments. Due to mathematical complexity, we restrict us to evaluate the memory function $M(z)$ defined in eqn.(\ref{mem2}) at $n=2$, i.e. by considering upto the $\ddot{J}-\ddot{J}$ correlation. We proceed  as follows. We begin with the evaluation of $\langle\langle \ddot{J};\ddot{J}\rangle\rangle_{z}$, which is defined as,
\begin{eqnarray} 
\langle\langle \ddot{J};\ddot{J}\rangle\rangle_{z} &=& -\langle\langle [\dot{J},H];[\dot{J},H]\rangle\rangle_{z} \nonumber\\
&=& \langle\langle \left[[J,H],H\right];\left[[J,H],H\right]\rangle\rangle_{z}.
\end{eqnarray}
Now considering the non-interacting and the interacting parts of the Hamiltonian separately the above equation can be rewritten as, 
\begin{eqnarray} 
\langle\langle \ddot{J};\ddot{J}\rangle\rangle_{z} &=& \langle\langle \left[[J,H_{\text{imp}}],H_{0}\right];
\left[[J,H_{\text{imp}}],H_{0}\right]\rangle\rangle_{z} \nn
&& + \langle\langle \left[[J,H_{\text{imp}}],
H_{\text{imp}}\right];\left[[J,H_{\text{imp}}],H_{0}\right]\rangle\rangle_{z} \nn 
&& + \langle\langle \left[[J,H_{\text{imp}}],H_{0}\right];\left[[J,H_{\text{imp}}],H_{\text{imp}}\right]
\rangle\rangle_{z}\nn
&& + \langle\langle \left[[J,H_{\text{imp}}],H_{\text{imp}}\right];\left[[J,H_{\text{imp}}],
H_{\text{imp}}\right]\rangle\rangle_{z}.\nn
\end{eqnarray}
The second term in the above expression is equal to the third term but with an opposite sign, due to the properties of the commutators.  Hence they cancel each other and thus we obtain,
\begin{eqnarray}
\nonumber
\langle\langle \ddot{J};\ddot{J}\rangle\rangle_{z} &=&\langle\langle \left[[J,H_{\text{imp}}],H_{0}\right];
\left[[J,H_{\text{imp}}],H_{0}\right]\rangle\rangle_{z}\nn
&&  + \langle\langle \left[[J,H_{\text{imp}}],
H_{\text{imp}}\right];\left[[J,H_{\text{imp}}],H_{\text{imp}}\right]\rangle\rangle_{z}. \nn
\label{jdoubledot}
\end{eqnarray}
To find the exact expression for the left hand side of the above equation, calculations can be performed in a way similar to that of the $\langle\langle \dot{J};\dot{J}\rangle\rangle_{z}$ in section \ref{sec: Firstmoment}. The details of which are presented in appendix \ref{app:B}. After several algebraic manipulations, we obtain,
\begin{widetext}
\begin{eqnarray}
\langle\langle \ddot{J};\ddot{J}\rangle\rangle_{z}&=&  \frac{2}{3}  \frac{N_{\text{imp}}U^{2}m^{2}}{\pi^{4}} \int_{0}^{\infty} d\epsilon \int_{0}^{\infty} 
 d\epsilon' \sqrt{\epsilon\epsilon'} \left( \epsilon + \epsilon'\right) \left( \epsilon - \epsilon'\right)^{2} 
 \frac{f(\epsilon)-f(\epsilon')}{z+\epsilon-\epsilon'} \nn
&& + \frac{2}{3} \frac{(N_{imp}U^{2})^{2}m^{2}}{\pi^{4}} \int_{0}^{\infty} d\epsilon \int_{0}^{\infty} d\epsilon' 
\sqrt{\epsilon\epsilon'} \left(\epsilon + \epsilon' \right) \frac{f(\epsilon)-f(\epsilon')}{z+\epsilon-\epsilon'}.
\label{jdoubledot4}
\end{eqnarray}
Using eqn.(\ref{jdoubledot4}) and performing the energy integrals as done in the case of first moment (eqn.(\ref{jdot2})), in eqn.(\ref{mem2}), the expression for the memory function $M(z)$ becomes,
\begin{eqnarray} \nonumber
M(z) &=& \frac{2}{3} \frac{m^{3}}{\pi^{4}}\frac{1}{N_{e}} \left\lbrace -\frac{2}{z} N_{\text{imp}} U^{2} 
\int_{0}^{\infty} d\epsilon \int_{0}^{\infty} d\epsilon' \epsilon \sqrt{\epsilon\epsilon'} \frac{f(\epsilon)-
f(\epsilon')}{\epsilon-\epsilon'}\right. \\ \nonumber
&& \left.-\frac{1}{z^{2}} N_{\text{imp}} U^{2}\int_{0}^{\infty} d\epsilon \int_{0}^{\infty} d\epsilon' 
\sqrt{\epsilon\epsilon'} \left( \epsilon + \epsilon'\right) \left( \epsilon - \epsilon'\right)^{2} 
\frac{f(\epsilon)-f(\epsilon')}{(z+\epsilon-\epsilon')(\epsilon-\epsilon')}\right. \\
&& \left. -\frac{1}{z^{2}} (N_{\text{imp}} U^{2})^{2} \int_{0}^{\infty} d\epsilon \int_{0}^{\infty} 
d\epsilon' \sqrt{\epsilon\epsilon'} \left(\epsilon + \epsilon' \right) \frac{f(\epsilon)-f(\epsilon')}
{(z+\epsilon-\epsilon')(\epsilon-\epsilon')} \right\rbrace.
\end{eqnarray}
After further algebraic manipulations, the expression for the complex memory function $M(z)$ reduce to
\begin{eqnarray} \nonumber
M(z)&=& \frac{2}{3} \frac{m^{3}}{\pi^{4}}\frac{1}{N_{e}} \int_{0}^{\infty} d\epsilon \int_{0}^{\infty} d\epsilon'\sqrt{\epsilon\epsilon'} 
\frac{f(\epsilon)-f(\epsilon')}{\epsilon-\epsilon'} \left\lbrace -N_{\text{imp}} U^{2} \frac{\epsilon+\epsilon'}
{z+\epsilon-\epsilon'}\right. \\ 
&& \left. - (N_{\text{imp}} U^{2})^{2} \frac{\epsilon+\epsilon'}{(\epsilon-\epsilon')^{2}(z+\epsilon-\epsilon')} + 
\frac{2}{z} (N_{\text{imp}} U^{2})^{2}  \frac{\epsilon}{(\epsilon-\epsilon')^{2}} \right\rbrace.
\end{eqnarray}
We are interested in the frequency dependent character of imaginary part of memory function  $M''(\omega)$ as a function of real frequency. On performing analytic continuation, i.e. $z 
\rightarrow \omega +i\eta$, $\eta \rightarrow 0$ 
, the expression for $M''(\omega)$ becomes,
\begin{eqnarray} \nonumber
M''(\omega) &=&  \frac{2}{3} \frac{m^{3}}{\pi^{3}} \frac{1}{N_{e}} \int_{0}^{\infty} d\epsilon \int_{0}^{\infty} 
d\epsilon'\sqrt{\epsilon\epsilon'} \frac{f(\epsilon)-f(\epsilon')}{\epsilon-\epsilon'} \delta(\omega+ \epsilon 
-\epsilon')\\
&& \left\lbrace N_{\text{imp}} U^{2} (\epsilon+\epsilon') + (N_{\text{imp}} U^{2})^{2} \frac{\epsilon+
\epsilon'}{(\epsilon-\epsilon')^{2}} - 2(N_{\text{imp}} U^{2})^{2}  \frac{\epsilon}{(\epsilon-\epsilon')^{2}} \delta(\omega)  \right\rbrace.
\end{eqnarray}
Now performing one of the energy integral, i.e. the integral over $\epsilon'$, the above expression for the memory function at frequeny $\omega > 0$ reduces to,
\begin{eqnarray}
M''(\omega) &=&  \frac{2}{3} \frac{m^{3}}{\pi^{3}} \frac{1}{N_{e}} \int_{0}^{\infty} d\epsilon \sqrt{\epsilon
(\epsilon+\omega)} \frac{f(\epsilon)-f(\epsilon+\omega)}{\omega}  (2\epsilon+\omega) \left\lbrace N_{\text{imp}} 
U^{2} + (N_{\text{imp}} U^{2})^{2} \frac{1}{\omega^{2}} \right\rbrace.
\label{finalmem}
\end{eqnarray}
\end{widetext} 
This is an expression of imaginary part of the memory function for electrons in metal, within the second order truncation of our proposed moment expansion for correlation function. Here the first term within the braces corresponds to the contribution from the first moment\cite{gotze_72} and the second term is the contribution from the second moment to the memory function. The frequency dependent behavior of the above expression for the imaginary part of the memory function or the scattering rate with different interaction strength $U$, impurity $N_{\text{imp}}$ and $T$ is discussed in next section.
\section{Results and Comparison}
\label{sec: results}

Eqn.(\ref{finalmem}), describes the imaginary part of the memory function or the scattering rate as 
a function of $\omega$, $U$, $N_{\text{imp}}$ and $T$ within a second order in moment expansion. We compare it with the imaginary part of the memory function obtained in eqn.(\ref{gwmem}), within a first order in moment expansion\cite{gotze_72}. The validity of truncating such an expansion at the $n$-th order is valid when the $n$-th term in the expansion is smaller than the ($n-1$)-th term. In the present work we restrict us at the second order. In this case to check the validity of our results,  we define an energy scale $\omega_{0}$ above which the present high frequency expansion is valid. By taking the ratio of second order term to the first order term, the condition becomes $\frac{1}{\omega^2} \frac{\langle \ddot{J} \vert \ddot{J} \rangle}{\langle \dot{J} \vert \dot{J} \rangle}<< 1$. From eqn.(\ref{finalmem}), the above criterion translates to $\frac{N_{\text{imp}}U^2}{\omega^2}<<1$.  This implies that our results are valid  if  the condition $\omega \geq (N_{\text{imp}}U^2)^{1/2} (= \omega_{0})$ is satisfied.\\
In fig.\ref{fig: comparison1}, we plot normalized imaginary part of MF $M''(\omega)/M_{0}$ as a function of frequency $\omega$ for both the cases (upto the first moment and the second moment), keeping other parameters fixed. 
\figcomp
\figcompN
\figcompU \\
In fig.\ref{fig: comp_moment_T10}, the scattering rates are shown at temperature $T=10$K. It is observed that at high frequency regime, the  result which includes the second moment contribution agrees well with the previous result (which includes only the first moment)\cite{gotze_72}. But above the defined energy scale $\omega_{0}$ (which is $0.004$ in this figure), results deviate from each other. The second moment contributes more in the later deviation and thus increasing the magnitude of the scattering rate compared to the case with only the first moment. 
We see that the magnitude of the scattering rate in this case is high as compared to the case with $n=1$ term of $M''(\omega)$. Similarly, the scattering rates are plotted at a different temperature $T=200$K in fig.\ref{fig: comp_moment_T200}. Here we observe the 
same behavior as in the previous figure, with temperature induced enhancement in the magnitude of the scattering rates. 

In fig.\ref{fig: comparison2}, again we plot the scattering rates fixing the temperature for different impurity densities $N_{\text{imp}} = 0.01$ and $0.04$. We observe the same trend in both cases similar to the previous figure. Here the increase in the impurity density increases the scattering centers which leads to higher magnitude to the scattering rates. Also, here the results are valid for frequency greater than $0.01$ and $0.02$ in figures \ref{fig: comp_moment_Nimpp01} and \ref{fig: comp_moment_Nimpp04} respectively.
From both the figures \ref{fig: comparison1} and \ref{fig: comparison2} we find that the scattering rate with the first moment approximation is valid only for high frequency regime and the truncation becomes more severe as one increases the interaction strength.

To elaborate its dependence on the interaction strength $U$, the plot of the scattering rate with $U$ at fixed frequency, $N_{\text{imp}}$ and temperature is shown in fig.\ref{fig: comp_moment_interaction}. In fig.\ref{fig: u1}, the scattering rate is shown at a small frequency $\omega=0.02$eV at which earlier we see that there is deviation in the results of memory function with different moment expansions. Here we find that the increase of $U$ increases the scattering rate at low frequency due to the presence of the term $(N_{\text{imp}}U^2)^{2}$ in the moment expansion of the memory function. In fig.\ref{fig: u2} we observe that at a higher frequency ($\omega=0.2$eV), difference in $M''(\omega)$ with the increase of interaction strength, from two approximations becomes insignificant. More discussions on these results are presented in the next section. 

\section{Discussion}
\label{sec: discussion}
It is often convenient to express a frequency dependent response function in terms of a memory function or ``multi-particle self energy''\cite{nabyendu_16}. In this work we propose a  series expansion for the memory function for optical conductivity or the current current correlation function. We show that, many of the previous works\cite{gotze_72, arfi_92, nabyendu_15, bhalla_15}, which address the optical conductivity of the metals within the memory function formalism, are equivalent to restricting at the lowest order in this expansion. We perform a higher order calculation for the same in the presence of electron-impurity interactions and compare our results with the results from one of the celebrated previous work\cite{gotze_72}. In all these approaches, one needs to calculate the current-current correlation function ($\langle JJ \rangle$), a two particle correlator with some approximations. In summary, conventional Kubo approach\cite{mahan_book} decouples $\langle J J \rangle$ correlation into a product of single particle correlators whereas G\"otze-W\"olfle\cite{gotze_72} first writes it in terms of $\langle \dot{J} \dot{J} \rangle$ and then use single particle decoupling. In the present approach, we extend the later work further and write $\langle J J \rangle$ in terms of $\langle \dot{J} \dot{J} \rangle$ and $\langle \ddot{J} \ddot{J} \rangle$ and use single particle decoupling of $\langle \ddot{J} \ddot{J} \rangle$. We see large discrepancy between the two results from the two approaches in the low frequency regime and also for higher impurity strengths.

These results are in accord with our proposal and also physically sensible. If we look at our expansion (eqn. \ref{mem2}), we see that as we go to the higher frequencies, the contributions from the higher order moments become more and more irrelevant. On the other hand, higher time derivatives of the current operator involves the higher power of impurity strengths. Thus the inclusion of the higher moments is equivalent to including higher order contribution in the perturbation theory. Inclusion of the effects from higher moments are also manifested in fig.\ref{fig: comp_moment_interaction}  where variation of the scattering rates at a certain frequency with the impurity strengths are shown. In this figure we see that the scattering rate is increasing with impurity strength and the inclusion of higher order contribution leads to higher scattering rates. The results at very low frequency $(\w << \omega_0)$ should not be trusted much. As discussed earlier, in this regime the present approximation is not valid. In case when $\omega_0$ is sufficiently small, result from the present method can be trusted even upto lower frequency. But we see that the second moment contribution to the memory function is $M_{2}''\sim \omega_0^2/\omega^2 M_{1}''$. This implies that the results for the memory function are in accord with the condition $\vert M(z) \vert << \vert z \vert$. This scenario can be clearly seen in the fig.\ref{fig: comparison1} and \ref{fig: comparison2} where the memory function $M''(\omega)$ is of very small magnitude as compared to the frequency $\omega$.
  
To summarize, our proposal is mathematically simpler compared to the previous attempts \cite{plakida_96, plakida_97} to calculate the memory function for the electronic conductivity beyond the lowest order perturbative calculations\cite{gotze_72}. Within this systematic expansion, we can include interaction effects up-to required order depending on its strength. This method in principle, can be applied for metals with other interactions as well as for non-metallic electronic systems\cite{subir_14, subir_15} to estimate higher order perturbative corrections.
\appendix
\section{Calculation of $\langle J \vert \dot{J} \rangle$}
\label{app: A}
Consider that the ensemble average of current operators at same time argument is represented by\\
\begin{equation}
\langle J \vert J \rangle = C
\end{equation}
where $C$ is some constant. \\
Now, differentiate above equation w.r.t. time
\begin{eqnarray} \nonumber
\langle \dot{J} \vert J \rangle + \langle J \vert \dot{J} \rangle &=& 0 \\
\langle \dot{J} \vert J \rangle &=& - \langle J \vert \dot{J} \rangle.
\label{eqn: JJdot1}
\end{eqnarray}
In another way, the ensemble average of $J$ and $\dot{J}$ can be expressed as
\begin{eqnarray} \nonumber
\langle \dot{J} \vert J \rangle &=& \text{tr} (\rho [H, J] J) \\ \nonumber
&=& \text{tr} (\rho HJJ) - \text{tr} (\rho JHJ)\\ \nonumber
&=& \text{tr} (\rho J[H, J])\\ 
&=& \langle J \vert \dot{J} \rangle.
\label{eqn: JJdot2}
\end{eqnarray}
From equations (\ref{eqn: JJdot1}) and (\ref{eqn: JJdot2}), we conclude that $\langle J \vert \dot{J} \rangle = 0$.
\section{Detailed calculation of the higher order contribution}
\label{app:B}
\begin{widetext}
To calculate $\langle\langle \ddot{J};\ddot{J}\rangle\rangle_{z}$ we first calculate the first term of eqn.(\ref{jdoubledot}). For this we need $\left[[J,H_{\text{imp}}],H_{0}\right]$ which using eqn.(\ref{unperturbed}) and (\ref{firstj}) becomes,

\begin{eqnarray} 
\left[[J,H_{\text{imp}}],H_{0}\right] &=& \frac{1}{N} \sum_{j,\textbf{k},\textbf{k}'} \langle \textbf{k} 
\vert U^{j} \vert \textbf{k}' \rangle \left( v_{x}(\textbf{k})-v_{x}(\textbf{k}')\right) \left( \epsilon_{\textbf{k}'} - 
\epsilon_{\textbf{k}} \right) c^{\dagger}_{\textbf{k}}c_{\textbf{k}'}.
\end{eqnarray}
Using the above expression, the first term of eqn.(\ref{jdoubledot}) becomes
\begin{eqnarray} 
&=& \frac{1}{N^{2}} \sum_{j,\textbf{k},\textbf{k}'} \sum_{i,\textbf{p},\textbf{p}'} \langle \textbf{k} \vert U^{j} 
\vert \textbf{k}' \rangle \langle \textbf{p} \vert U^{i} \vert \textbf{p}' \rangle \left( v_{x}(\textbf{k})-v_{x}
(\textbf{k}')\right)  \left( v_{x}(\textbf{p})-v_{x}(\textbf{p}')\right)\left( \epsilon_{\textbf{k}'} - 
\epsilon_{\textbf{k}} \right)\left( \epsilon_{\textbf{p}'} - \epsilon_{\textbf{p}} \right) \langle\langle 
c^{\dagger}_{\textbf{k}}c_{\textbf{k}'}; c^{\dagger}_{\textbf{p}}c_{\textbf{p}'}\rangle\rangle_{z}. 
\end{eqnarray}
Here again we will consider the case of $i=j$ as considered in eqn.(\ref{ijcase}) and using eqn.(\ref{chi1}) 
with performing time integration and ensemble average, the above equation reduces to
\begin{eqnarray} 
&=& \frac{2N_{\text{imp}}}{N^{2}} \sum_{\textbf{k},\textbf{k}'} \vert \langle \textbf{k} \vert U \vert 
\textbf{k}' \rangle  \vert^{2} \left( v_{x}(\textbf{k})-v_{x}(\textbf{k}')\right)^{2} \left( \epsilon_{\textbf{k}} - 
\epsilon_{\textbf{k}'} \right)^{2} \frac{ f(\textbf{k}) -f(\textbf{k}')}{z+\epsilon_{\textbf{k}} - \epsilon_{\textbf{k}'}}.
\end{eqnarray}
This expression is further simplified by converting summations into energy integrals and ignoring the momentum dependence of $U$ as
\begin{eqnarray} 
&=& \frac{2}{3} N_{\text{imp}} \frac{U^{2}m^{2}}{\pi^{4}} \int_{0}^{\infty} d\epsilon \int_{0}^{\infty} d\epsilon' 
\sqrt{\epsilon\epsilon'} \left( \epsilon + \epsilon'\right) \left( \epsilon - \epsilon'\right)^{2} 
\frac{f(\epsilon)-f(\epsilon')}{z+\epsilon-\epsilon'}.
\label{jdoubledot2}
\end{eqnarray}
Now we perform the calculations for the second term of eqn.(\ref{jdoubledot}). First,  $\left[[J,H_{\text{imp}}],
H_{\text{imp}}\right]$ using eqn.(\ref{imp}) and (\ref{firstj}) is written as
\begin{eqnarray} \nonumber
\left[[J,H_{\text{imp}}],H_{\text{imp}}\right] &=& \frac{1}{N^{2}} \sum_{j,\textbf{k},\textbf{k}'} 
\sum_{i,\textbf{p},\textbf{p}'} \langle \textbf{k} \vert U^{j} \vert \textbf{k}' \rangle \langle \textbf{p} \vert U^{i} 
\vert \textbf{p}' \rangle \left( v_{x}(\textbf{k})-v_{x}(\textbf{k}')\right) \left[c^{\dagger}_{\textbf{k}}c_{\textbf{k}'} , 
c^{\dagger}_{\textbf{p}}c_{\textbf{p}'} \right] \\ \nonumber
&=& \frac{N_{\text{imp}}}{N^{2}}\sum_{\textbf{k},\textbf{k}',\textbf{p}} \langle \textbf{k} \vert U \vert \textbf{k}' 
\rangle \langle \textbf{k}' \vert U \vert \textbf{p} \rangle \left( v_{x}(\textbf{k}) - 2v_{x}(\textbf{k}') + v_{x}
(\textbf{p}) \right) c^{\dagger}_{\textbf{k}}c_{\textbf{p}}. \\
\end{eqnarray}
Using this, $\langle\langle \left[[J,H_{\text{imp}}],H_{\text{imp}}\right];\left[[J,H_{\text{imp}}],H_{\text{imp}}\right]
\rangle\rangle_{z}$ can be written as
\begin{eqnarray} \nonumber
&=& 2 \frac{N^{2}_{imp}}{N^{4}} \sum_{\textbf{k},\textbf{k}',\textbf{p}}\sum_{\textbf{r},\textbf{r}',\textbf{l}} 
\langle \textbf{k} \vert U \vert \textbf{k}' \rangle \langle \textbf{k}' \vert U \vert \textbf{p} \rangle \langle 
\textbf{r} \vert U \vert \textbf{r}' \rangle \langle \textbf{r}' \vert U \vert \textbf{l} \rangle \left( 
v_{x}(\textbf{k}) - 2v_{x}(\textbf{k}') + v_{x}(\textbf{p}) \right)\left( v_{x}(\textbf{r}) - 2v_{x}(\textbf{r}') + 
v_{x}(\textbf{l}) \right) \langle\langle c^{\dagger}_{\textbf{k}}c_{\textbf{p}} ; c^{\dagger}_{\textbf{r}}c_{\textbf{l}} \rangle\rangle_{z}. \\
\label{jbig}
\end{eqnarray}
After calculating $\langle\langle c^{\dagger}_{\textbf{k}}c_{\textbf{p}} ; c^{\dagger}_{\textbf{r}}
c_{\textbf{l}} \rangle\rangle_{z}$ with help of eqn.(\ref{chi1}) and substituting in eqn.(\ref{jbig}) 
and taking $U$ as independent of momentum, $\langle\langle \left[[J,H_{\text{imp}}],H_{\text{imp}}\right];\left[[J,H_{\text{imp}}],H_{\text{imp}}\right]\rangle\rangle_{z}$ can be expressed as
\begin{eqnarray}
&=& 2 \frac{N^{2}_{imp}U^{4}}{N^{4}m^{2}}\sum_{\textbf{k},\textbf{k}',\textbf{p},\textbf{r}'}  
\frac{1}{z+\epsilon_{\textbf{k}}-\epsilon_{\textbf{p}}} \left( k_{x} - 2k'_{x} + p_{x} \right) 
\left( p_{x} - 2r'_{x} + k_{x} \right) \left( f_{\textbf{k}} - f_{\textbf{p}} \right).
\end{eqnarray}
After doing algebra, this above expression can be written as
\begin{eqnarray}
&=& \frac{2}{3} \frac{N^{2}_{imp}U^{4}m^{2}}{\pi^{4}} \int_{0}^{\infty} d\epsilon \int_{0}^{\infty} 
d\epsilon' \sqrt{\epsilon\epsilon'} \left(\epsilon + \epsilon' \right) \frac{f(\epsilon)-f(\epsilon')}{z+\epsilon-\epsilon'}.
\label{jdoubledot3}
\end{eqnarray}
Substituting eqns.(\ref{jdoubledot2}) and (\ref{jdoubledot3}) in eqn.(\ref{jdoubledot}), we have
\begin{eqnarray} \nonumber
\langle\langle \ddot{J} ; \ddot{J} \rangle\rangle_{z} &=&  \frac{2}{3}  \frac{N_{\text{imp}}U^{2}m^{2}}{\pi^{4}} \int_{0}^{\infty} d\epsilon \int_{0}^{\infty} 
 d\epsilon' \sqrt{\epsilon\epsilon'} \left( \epsilon + \epsilon'\right) \left( \epsilon - \epsilon'\right)^{2} 
 \frac{f(\epsilon)-f(\epsilon')}{z+\epsilon-\epsilon'} \\
&& + \frac{2}{3} \frac{N^{2}_{imp}U^{4}m^{2}}{\pi^{4}} \int_{0}^{\infty} d\epsilon \int_{0}^{\infty} d\epsilon' 
\sqrt{\epsilon\epsilon'} \left(\epsilon + \epsilon' \right) \frac{f(\epsilon)-f(\epsilon')}{z+\epsilon-\epsilon'}.
\end{eqnarray}
\end{widetext}

\end{document}